\documentclass[12pt]{article}
\usepackage{graphicx}
\usepackage{amsmath}

\oddsidemargin  - 5 mm
\evensidemargin - 5 mm
\newtheorem{theorem}{Theorem}

\newtheorem{proposition}[theorem]{Proposition}

\input{tcilatex}

\begin{document}

\date{}
\title{Elliptic Curves and Algebraic Geometry Approach in Gravity Theory
I.The General Approach }
\author{Bogdan G. Dimitrov \thanks{%
Electronic mail: bogdan@theor.jinr.ru} \\
Bogoliubov Laboratory for Theoretical Physics\\
Joint Institute for Nuclear Research \\
6 Joliot-Curie str. \\
Dubna 141 980, Russia}
\maketitle

\begin{abstract}
\ \ 

\ \ \ Based on the distinction between the covariant and contravariant
metric tensor components in the framework of the affine geometry approach
and also on the choice of the contravariant components, it was shown that a
wide variety of third, fourth, fifth, sixth, seventh - degree algebraic
equations exists in gravity theory. This fact, together with the derivation
of the algebraic equations for a generally defined contravariant tensor
components in this paper, are important in view of finding new solutions of
the Einstein's equations, if they are treated as algebraic ones. Some
important properties of the introduced in hep-th/0107231 more general
connection have been also proved - it possesses affine transformation
properties and \ it is an equiaffine one. Basic and important knowledge
about the affine geometry approach and about gravitational theories with
covariant and contravariant connections and metrics is also given with the
purpose of demonstrating when and how these theories can be related to the
proposed algebraic approach and to the existing theory of gravity and
relativistic hydrodynamics. 
\end{abstract}

\section{\protect\bigskip \protect\bigskip INTRODUCTION}

Inhomogeneous cosmological models have been intensively studied in the past
in reference to colliding gravitational \bigskip waves [1] or singularity
structure and generalizations of the Bondi - Tolman and Eardley-Liang-Sachs
metrics [2, 3]. In these models the inhomogeneous metric is assumed to be of
the form [2] 
\begin{equation}
ds^{2}=dt^{2}-e^{2\alpha (t,r,y,z)}dr^{2}-e^{2\beta (t,r,y,z)}(dy^{2}+dz^{2})
\tag{1.1}
\end{equation}%
(or with $r\rightarrow z$ and $z\rightarrow x$) with an energy-momentum
tensor $G_{\mu \nu }=k\rho u_{\mu }u_{\nu }$ for the irrotational dust. The
functions $\alpha (t,r,y,z)$ and $\beta (t,r,y,z)$, determined by the
Einstein's equations, are chosen in a special form [4], so that the
integrations of (some) of the components of the Einstein's equations is
ensured.

A\ nice feature of the approach is that in the limit $t\rightarrow \infty $
[5] and under a special choice of the pressure as a definite function of
time the metric approaches an isotropic form [4]. Other papers, also
following the approach of Szafron-Szekerez are [6,7]. In [7], after an
integration of one of the components - $G_{1}^{0}$ of the Einstein's
equations, a solution in terms of an elliptic function is obtained.

In different notations, but again in the framework of the Szafron-Szekerez
approach the same integrated in [7] nonlinear differential equation \ 
\begin{equation}
\left( \frac{\partial \Phi }{\partial t}\right) ^{2}=-K(z)+2M(z)\Phi ^{-1}+%
\frac{1}{3}\Lambda \Phi ^{2}  \tag{1.2}
\end{equation}%
was obtained in the paper [8] of Kraniotis and Whitehouse. They make the
useful observation that (1.2) in fact defines a (cubic) algebraic equation
for an elliptic curve, which according to the standard algebraic geometry
prescribtions (see [9] for an elementary, but comprehensive and contemporary
introduction) can be parametrized with the elliptic Weierstrass function 
\begin{equation}
\rho (z)=\frac{1}{z^{2}}+\sum\limits_{\omega }\left[ \frac{1}{(z-\omega )^{2}%
}-\frac{1}{\omega ^{2}}\right]  \tag{1.3}
\end{equation}%
and the summation is over the poles in the complex plane. Two important
problems immediately arise, which so far have remained without an answer%
\textbf{: }

1. The parametrization procedure with the elliptic Weierstrass function in
algebraic geometry is adjusted for cubic algebraic equations with number
coefficients! Unfortunately, equation (1.2) is not of this type, since it
has coefficient functions in front of the variable $\Phi $, which depend on
the complex variable $z$. In view of this, it makes no sense to define
\textquotedblright Weierstrass invariants\textquotedblright\ as 
\begin{equation}
g_{2}=\frac{K^{2}(z)}{12}\text{ \ \ ; \ \ \ }g_{3}=\frac{1}{216}K^{3}(z)-%
\frac{1}{12}\Lambda M^{2}(z)\text{ \ \ ,}  \tag{1.4}
\end{equation}%
since the above functions have to be set up equal to the complex numbers $%
g_{2}$ and $g_{3}$ (the s. c. Eisenstein series) 
\begin{equation}
g_{2}=60\sum\limits_{\omega \subset \Gamma }\frac{1}{\omega ^{4}}%
=\sum\limits_{n,m}\frac{1}{(n+m\tau )^{4}}\text{ \ \ \ ,}  \tag{1.5}
\end{equation}%
\begin{equation}
g_{3}=140\sum\limits_{\omega \subset \Gamma }\frac{1}{\omega ^{6}}%
=\sum\limits_{n,m}\frac{1}{(n+m\tau )^{6}}\text{ \ \ \ \ }  \tag{1.6}
\end{equation}%
and therefore additional equations\textbf{\ }have to be satisfied in order
to ensure the parametrization with the Weierstrass function.

2. Is the Szekerez - Szafron metric the only case, when the parametrization
with the Weierstrass function is possible? Closely related to this problem
is the following one - is only one of the components of the Einstein's
equation parametrizable with \ $\rho (z)$ and its derivative?

This series of three papers has the aim to present an adequate mathematical
algorithm for finding solutions of the Einstein's equations in terms of
elliptic functions. This approach is based on the clear distinction between
covariant and contravariant metric tensor components within the s.c affine
geometry approach, which will be clarified further in Section 2. Afterwords,
a cubic algebraic equation in terms of the contravariant metric components
will be obtained, which according to the general prescription and the
algorithm in the previous paper [10] can be parametrized with the
Weierstrass function and its derivative. Respectively, if the contravariant
components are assumed to be known, then a cubic (or a quartic) algebraic
equation with respect to the covariant components can be investigated and
parametrized again with the Weierstrass function. \textit{Thus it will turn
out that the parametrization with the Weierstrass function will be possible
not only in the Szafron-Szekeres case, but also in the general case due to
the "cubic" algebraic structure of the gravitational Lagrangian.} This is an
important point since valuable cosmological characteristics for
observational cosmology such as the Hubble's constant $H(t)=\frac{\overset{.}%
{R}(t)}{R(t)}$ and the deceleration parameter $q=-\frac{\overset{..}{R}%
(t)R(t)}{\overset{.}{R}^{2}(t)}$ may be expressed in terms of the Jacobi's
theta function and of the Weierstrass elliptic function respectively [8].
Unfortunately, in the paper [8] the Eisenstein series (1.5-1.6) have not
been taken into account, due to which the obtained expression for the metric
will be another one and will be modified.

Instead of searching elliptic solutions of the Einstein's equations for each
separate case of a given metric, as in nearly all of the mentioned papers,
in this series of papers another approach will be proposed. First, a cubic
algebraic equataion will be parametrized with respect to one of the
contravariant components, following the approach in a previous paper [10].
In the second part, this parametrization will be extended to more than one
variable in the \textit{multivariable cubic algebraic equation.} This will
be a substantial and new development, different from the standard algebraic
geometry approach, in which \textit{only two-dimensional cubic equations }%
are parametrized with the (elliptic) Weierstrass function and its
derivative. Finally, in the third part the dependence of the generalized
coordinates $X^{i}=X^{i}(x_{1},x_{2},x_{3},.....,x_{n})$ on the complex
variable $z$ will be established from a derived \textit{system of
first-order nonlinear differential equations.} The generalized coordinates
can be regarded as $n-$ dimensional hypersurfaces, defining a transition
from an initially defined set of coordinates $x_{1},x_{2},x_{3},.....,x_{n}$
on a chosen manifold to another set of the generalized coordinates $%
X^{1},X^{2},.....,X^{n}$. Since the covariant metric components $g_{ij}$
also depend on these coordinates, this means that their dependence on the
complex variable $z$ will also be known. In other words, at the end of the
applied approach, each initially given function $g_{ij}(t,\mathbf{x})$ of
the time and space coordinates will be expressed also as $g_{ij}(z)$. The
algebraic approach will be applied to the s .c. \textit{cubic algebraic
equation for reparametrization invariance of the gravitational Lagrangian},
but further it will be shown that not only the approach will be applicable
in the general case of an arbitrary contravariant tensor, but also concrete
solutions for the metric $g_{ij}(z)$ will be given in the case of specially
chosen simple metrics.

The first part of the present paper continues and develops further the
approach from the previous paper [10], where a definite choice of the
contravariant metric tensor was made in the form of the factorized product $%
\widetilde{g}^{ij}=dX^{i}dX^{j}$. The differentials $dX^{i}$ are assumed to
lie in the tangent space $T_{X}$ of the generalized coordinates. In Section
2 of the present paper some basic facts about the affine geometry approach
and the s.c. \textit{gravitational theory with covariant and contravariant
metrics and connections (GTCCMC)} will be reminded, but also  some new
material, related to relativistic hydrodynamics in the context of GTCCMC is
added. The basic and very important idea in this section is to show that
GTCCMC are already "incorporated" in the theoretical framework of the
already known gravitational theories - as an example the known projective
formalism is taken, but at the same time in certain theories (such as the
Arnowitt-Deser-Misner $3+1$ decomposition), certain assumptions are made so
that they do not fall within the class of GTCCMC. This is an interesting
observation, since it clearly shows that the limiting assumptions can
naturally be removed. In the next Section 3 it will be demonstrated briefly
how the cubic algebraic equation with respect to the differentials $dX^{i}$
was derived in [10], but in fact the aim will be to show that depending on
the choice of variables in the gravitational Lagrangian or in the Einstein's
equations, a \textit{wide variety of algebraic equations} (of third, fourth,
fifth, seventh, tenth- degree) in gravity theory may be treated, if a 
\textit{distinction between the covariant metric tensor components and the
contravariant ones is made.} This idea, originally set up by Schouten and
Schmutzer, was further developed in the papers \ [13, 14]  mainly with the
purpose of classification of such more general GTCCMC [13, 14]. Also, in
Section 3 the important and new physical notion of a \textit{"tensor length
scale"} is introduced in a natural way within the GTCCMC, and this notion is
a generalization of the metrical (scale) function $%
l(x)=ds^{2}=g_{ij}dx^{i}dx^{j}$ in usual gravity theory. In Section 4
intersecting algebraic varieties will be proposed as a method for obtaining
the  known solutions in the standard gravity theory. In Section 5 it will be
shown that the previously investigated in [10] under some restrictive
assumptions cubic algebraic equation for reparametrization invariance of the
gravitational Lagrangian and  the Einstein's equations can  be written as
algebraic ones also in the general case of an arbitrary contravariant tensor 
$g^{ij}$.    

The physical idea, which will be exploited in this paper will be: \textit{%
can such a gravitational theory with a more general contravariant tensor be
equivalent to the usual and known to us theory with a contravariant metric
tensor, which is at the same time the inverse one of the covariant one?} By 
\textit{\textquotedblright equivalence\textquotedblright }\ it is meant that
the gravitational Lagrangian in both approaches should be equal, on the base
of which the s.c. cubic algebraic equation for reparametrization invariance
(of the gravitational Lagrangian) was obtained in [10]. The derivation was
based also on the construction of another connection $\widetilde{\Gamma }%
_{kl}^{s}\equiv \frac{1}{2}dX^{i}dX^{s}(g_{ik,l}+g_{il,k}-g_{kl,i})$, which
is with contravariant tensor component replaced with the factorized product $%
dX^{i}dX^{s}$. The connection $\widetilde{\Gamma }_{kl}^{s}$ has two very
useful properties: 1. It may have an affine transformation law under a broad
variety of coordinate transformations (see Append. A2), which can be found
after solving a system of nonlinear differential equations. 2. $\widetilde{%
\Gamma }_{kl}^{s}$ is an equiaffine connection (see \ also Appendix A3 for
the elementary proof), which is a typical notion, introduced in classical
affine geometry [15, 16] and meaning that there exists a volume element,
which is preserved under a parallel displacement of a basic $n-$dimensional
vector $e\equiv e_{i_{1}i_{2}....i_{n}}$. Equivalently defined, $\widetilde{%
\Gamma }_{kl}^{s}$ is an equiaffine connection [15, 16] if it can be
represented in the form $\widetilde{\Gamma }_{ks}^{s}=\partial _{k}lge$,
where $e$ is a scalar quantity. This notion turns out to be very convenient
and important, since for such types of connections we can use the known
formulae for the Ricci tensor, but with the connection $\widetilde{\Gamma }%
_{kl}^{s}$ instead of the usual Christoffell one $\Gamma _{kl}^{s}$.
Moreover, the Ricci tensor $\widetilde{R}_{ij}$ will again be a symmetric
one, i.e. $\widetilde{R}_{ij}=\widetilde{R}_{ji}=\partial _{k}\widetilde{%
\Gamma }_{ij}^{k}-\partial _{i}\widetilde{\Gamma }_{kj}^{k}+\widetilde{%
\Gamma }_{kl}^{k}\widetilde{\Gamma }_{ij}^{l}-\widetilde{\Gamma }_{ki}^{m}%
\widetilde{\Gamma }_{jm}^{k}$.

In usual gravity theory, the contravariant components are at the same time
inverse to the covariant ones , and thus the correspondence between
\textquotedblright covectors\textquotedblright\ (in our terminology - these
are the \textquotedblright vectors\textquotedblright ) and the
\textquotedblright vectors\textquotedblright\ (i.e. the contravariant
vectors\textquotedblright ) is being set up, respectively there is
correspondence between covariant and contravariant tensors. By
"correspondence" it is meant that both these kinds of tensors satisfy the
matrix equation $g_{ij}g^{jk}=\delta _{i}^{k}$. However, within the
framework of affine geometry, such a correspondence is not necessarily to be
established (see again [15-18]) and both tensors have to be treated as 
\textit{different mathematical objects,} defined on one and the same
manifold. If both components constitute the algebraic variety, satisfying
the Einstein's equations, considered as a set of intersecting multivariable
cubic and quartic algebraic surfaces\textbf{\ (}further instead of cubic
surfaces we shall continue to use the terminology \textquotedblright cubic
curves\textquotedblright ), then one can speak about separate classes\textbf{%
\ }of solutions for the covariant metric tensor components and for the
contravariant ones.

If one assumes the existence of inverse contravariant metric tensor
components $\widetilde{g}^{jk}$ and considers the quadratic system $g_{ij}%
\widetilde{g}^{jk}=g_{ij}dX^{j}dX^{k}=\delta _{i}^{k}$ as intersecting with
the set of cubic and quartic algebraic Einstein's equations, then it might
be expected that the standardly known solutions of the Einstein's equations
should be recovered. However, this is not yet mathematically proved, neither
has this been formulated as a problem. General theorems for \ intersection
of \ algebraic curves of different (arbitrary)\ degrees are given in [19,
21, 22].

\section{\ AFFINE GEOMETRY APPROACH \ AND\ GRAVITATIONAL\ \ THEORIES\ \
WITH\ \ COVARIANT\ \ AND\ CONTRAVARIANT CONNECTIONS \ AND\ METRICS}

This section has the purpose to review some of the basic aspects of \textit{%
gravitational theories with covariant and contravariant metrics and
connections (GTCCMC)}, which would further allow the application of
algebraic geometry and of the theory of algebraic equations in gravity
theory. The section contains also some new material, concerning the
application of GTCCMC in relativistic hydrodynamics.  

It is known in gravity theory that the knowledge of the metric tensor $%
g_{ij} $ determines the space - time geometry, which means that the
Christoffell connection 
\begin{equation}
\Gamma _{ik}^{l}\equiv \frac{1}{2}g^{ls}(g_{ks,i}+g_{is,k}-g_{ik,s}) 
\tag{2.1}
\end{equation}%
and the Ricci tensor 
\begin{equation}
R_{ik}=\frac{\partial \Gamma _{ik}^{l}}{\partial x^{l}}-\frac{\partial
\Gamma _{il}^{l}}{\partial x^{k}}+\Gamma _{ik}^{l}\Gamma _{lm}^{m}-\Gamma
_{il}^{m}\Gamma _{km}^{l}\text{ \ \ \ }  \tag{2.2}
\end{equation}%
can be calculated.

It is useful to remember also from standard textbooks [24] the s. c.
Christoffell connection of the first kind: 
\begin{equation}
\Gamma _{i;kl}\equiv g_{im}\Gamma _{kl}^{m}=\frac{1}{2}%
(g_{ik,l}+g_{il,k}-g_{kl,i})\text{ \ ,}  \tag{2.3}
\end{equation}%
obtained from the expression for the zero covariant derivative $0=\nabla
_{l}g_{ik}=g_{ik,l}-g_{m(i}\Gamma _{k)l}^{m}$ $.$ By contraction of (2.3)
with another contravariant tensor field $\widetilde{g}^{is}$, one might as
well define another connection\textbf{:} 
\begin{equation}
\widetilde{\Gamma }_{kl}^{s}\equiv \widetilde{g}^{is}\Gamma _{i;kl}=%
\widetilde{g}^{is}g_{im}\Gamma _{kl}^{m}=\frac{1}{2}\widetilde{g}%
^{is}(g_{ik,l}+g_{il,k}-g_{kl,i})\text{ \ ,}  \tag{2.4}
\end{equation}%
not consistent with the initial metric $g_{ij}$. Clearly the connection
(2.4) is defined under the assumption that the contravariant metric tensor
components $\widetilde{g}^{is}$ are not to be considered to be the inverse
ones to the covariant components $g_{ij}$ and therefore $\widetilde{g}%
^{is}g_{im}\equiv f_{m}^{s}(\mathbf{x})$.

In fact, the definition $\widetilde{g}^{is}g_{im}\equiv f_{m}^{s}$ turns out
to be inherent to gravitational physics. For example, in the projective
formalism one decomposes the standardly defined metric tensor (with $%
g_{ij}g^{jk}=\delta _{i}^{k}$) as 
\begin{equation}
g_{ij}=p_{ij}+h_{ij}\text{ \ \ ,}  \tag{2.5}
\end{equation}%
together with the additional assumption that the two subspaces, on which the
projective tensor $p_{ij}$ and the tensor $h_{ij}$ are defined, are
orthogonal. This means that 
\begin{equation}
p_{ij}h^{jk}=0\text{ \ \ \ .}  \tag{2.6}
\end{equation}%
As a consequence 
\begin{equation}
p_{ij}p^{jk}=\delta _{i}^{k}-h_{ij}h^{jk}\neq \delta _{i}^{k}\text{ \ \ ,} 
\tag{2.7 }
\end{equation}
meaning that the contravariant projective metric components $p^{jk}$ are no
longer inverse to the covariant ones $p_{ij}$.

An example of gravitational theories with more than one connection are the
so called \textit{theories with affine connections and metrics} [13], in
which there is one connection $\Gamma _{\alpha \beta }^{\gamma }$ for the
case of a parallel transport of covariant basic vectors $\nabla _{e_{\beta
}}e_{\alpha }=\Gamma _{\alpha \beta }^{\gamma }$ $e_{\gamma }$ and a \textit{%
separate} connection $P_{\alpha \beta }^{\gamma }$ for the contravariant
basic vector $e^{\gamma }$, the defining equation for which is $\nabla
_{e_{\beta }}e^{\alpha }=P_{\gamma \beta }^{\alpha }$ $e^{\gamma }$. In
these theories, the contravariant vector and tensor fields are assumed to be 
\textit{not the inverse ones} to the covariant vector and tensor fields.
This implies that 
\begin{equation}
e_{\alpha }e^{\beta }\equiv f_{\alpha }^{\beta }(x)\neq \delta _{\alpha
}^{\beta }  \tag{2.8}
\end{equation}%
for such theories and consequently, a distinction is made between covariant
and contravariant metric tensors (and vectors too). Clearly, in the above
given case (2.7) of projective gravity, this theory should be considered as
a GTCCMC. In the same spirit, since the well - known Arnowitt - Deser -
Misner (ADM)\ (3+1) decomposition of spacetime [44, 45] is built upon the
projective transformation (2.5), it might be thought that it should also be
considered as such a theory. But in fact, the ADM (3+1) formalism definitely
is not an example for this, because \emph{due to the special identification
\ of the vector field's components [44, 45] with certain components of the
projective tensor }%
\begin{equation}
g_{00}:=-(N^{2}-N_{i}N^{i})\text{ \ ; \ }g^{00}:=-\frac{1}{N^{2}}\text{ \ ,}
\tag{2.9}
\end{equation}%
\begin{equation}
g_{ij}:=p_{ij}\text{ \ ; \ }g^{ij}:=p^{ij}-\frac{N^{i}N^{j}}{N^{2}}\text{ \
\ ,}  \tag{2.10}
\end{equation}%
\begin{equation}
g_{0i}:=N_{i}\text{ \ ; \ \ }g^{0i}:=\frac{N^{i}}{N^{2}}\text{ \ \ ,} 
\tag{2.11}
\end{equation}%
all the contravariant projective tensor components $p^{\alpha \beta }$ ($%
\alpha ,\beta ,\gamma =0,1,2,3$ ; $\ i,j=1,2,3$) \textit{turn out to be the
inverse ones} to the covariant projective components $p_{\alpha \gamma }$.
Indeed, it follows that 
\begin{equation}
p_{ij}p^{jk}=\delta _{i}^{k}\text{ \ \ \ ; \ \ }N_{i}N^{i}=N^{2}\text{ \ ; \ 
}N_{i}N^{j}=\delta _{i}^{j}\text{\ \ \ .}  \tag{2.12}
\end{equation}%
In the case of the ADM (3+1) decomposition, such an identification is indeed
possible and justified, since the gravitational field posseses coordinate
invariance, allowing to disentangle the dynamical degrees of freedom from
the gauge ones. But in the case when the tensors $h_{ij}$ are related with
some moving matter (with a prescribed motion) and an observer, "attached" to
this matter "measures" all the gravitational phenomena in his reference
system by means of the projective metric $p_{ij}$, this will be no longer
possible. Then the relation (2.7) will hold, and the resulting theory will
be \textit{a gravitational theory with covariant and contravariant metrics
and connections (GTCCMC).} Naturally, if the tensor $h_{ij}$ in (2.5) and
(2.7) is taken in the form $h_{ij}=\frac{1}{e}u_{i}u^{j}$ and if the vector
field $u$ (tangent at each point of the trajectory of the moving matter) is
assumed to be non-normalized (i.e. $e(x)=u_{i}u^{i}\neq 1$), then one would
have to work \textit{not within} the standard relativistic hydrodynamics
theory (where $p_{ij}=g_{ij}-u_{i}u_{j}$ and $p_{ij}p^{jk}=\delta
_{i}^{k}-u_{i}u^{k}$), but within the formalism of GTCCMC (where $%
p_{ij}p^{jk}=f_{i}^{k}=\delta _{i}^{k}-\frac{1}{e}u_{i}u^{k}\neq \delta
_{i}^{k}$).\ One may wonder why this should be so, since the last two
formulaes for $p_{ij}p^{jk}$ for both cases look very much alike, with the
exception of the "normalization" function $\frac{1}{e}$ in the second
formulae. But in what follows it shall become clear that in the \textit{%
first case} the right-hand side has a \textit{tensor transformation property}%
, while in the \textit{second case} due to the function $\frac{1}{e}$ there
would be no such property. And this shall turn out to be crucial.

In order to understand this also from another point of view, let us perform
a covariant differentiation of both sides of the relation (2.8). Then one
can obtain that the two connections are related in the following way [13] 
\begin{equation}
f_{j,k}^{i}=\Gamma _{jk}^{l}\text{ }f_{l}^{i}+P_{lk}^{i}f_{j}^{l}\text{ \ \
\ \ \ ;\ \ \ \ \ \ (}f_{j,k}^{i}=\partial _{k}f_{j}^{i}\text{) \ \ \ \ . } 
\tag{2.13}
\end{equation}

Note also the following important moment - $f_{\alpha }^{\beta }(x)$ are
considered to be the components of a \textit{function. }Otherwise, if they
are considered to be a (mixed) tensor quantity, the covariant
differentiation of the mixed tensor $f_{\alpha }^{\beta }(x)$ in the
right-hand side of $e_{\alpha }e^{\beta }\equiv f_{\alpha }^{\beta }(x)$
would give exactly the same expression as in the left-hand side. This would
mean that from a mathematical point of view there would be no justification
for the introduction of the second covariant connection $P_{lk}^{i}$ - the
covariant differentiation would give a quantity on the left-hand side, which
would be identically equal to $\nabla _{\gamma }f_{\alpha }^{\beta }(x)$ for 
\textit{every choice} of the two connections $\Gamma _{jk}^{l}$ and $%
P_{lk}^{i}$, including also for the standard case (Einsteinian gravity) $%
P_{lk}^{i}=-\Gamma _{jk}^{l}$. However, in view of the fact that $f_{\alpha
}^{\beta }(x)$ are related with the description of some moving matter in the
Universe, then a tensor transformation law should not be prescribed to them.
So they should remain components of a function and consequently, the
introduction of the second connection $P_{lk}^{i}$ is inevitable.

In confirmation of this, it can easily be seen that the quantity $\delta
_{i}^{k}-h_{ij}h^{jk}\neq \delta _{i}^{k}$ in (2.7), which is to be set up
equal to $f_{i}^{k}(x)$, \textit{does not} have a tensor transformation
property for \textit{arbitrarily} chosen tensor fields $h_{ij}$. More
concretely, it would have such a transformation property if the equality 
\begin{equation*}
\left( \delta _{i}^{j}-\frac{1}{e}h_{ik}h^{kj}\right) ^{^{\prime }}(\mathbf{X%
})=
\end{equation*}%
\begin{equation}
=\frac{\partial x^{\alpha }}{\partial X^{i}}\frac{\partial X^{j}}{\partial
x^{\beta }}\left( \delta _{\alpha }^{\beta }-\frac{1}{e}h_{\alpha \gamma
}h^{\gamma \beta }\right) (\mathbf{x})  \tag{2.14}
\end{equation}%
holds. Now since\textit{\ }$h_{ik}h^{kj}$ transforms as a tensor, then the
fulfillment of (2.14) would mean that the equality 
\begin{equation}
\delta _{i}^{j}=\frac{\partial x^{\alpha }}{\partial X^{i}}\frac{\partial
X^{j}}{\partial x^{\beta }}\delta _{\alpha }^{\beta }  \tag{2.15}
\end{equation}%
\ should hold for \textit{any} derivatives $\frac{\partial x^{\alpha }}{%
\partial X^{i}}$ and $\frac{\partial X^{j}}{\partial x^{\beta }}$. But if $%
t_{i}^{\alpha }:=$\ $\frac{\partial x^{\alpha }}{\partial X^{i}}$ and $%
t_{\beta }^{j}:=$\ $\frac{\partial X^{j}}{\partial x^{\beta }}$ are the
components of some set of tetrad fields, this would imply that this set is 
\textit{orthonormal}, i.e. $\delta _{i}^{j}=t_{i}^{\alpha }t_{\beta
}^{j}\delta _{\alpha }^{\beta }$ - a property, which now we shall prove to
be \textit{not consistent} with equality (2.7) $p_{ij}p^{jk}=\delta
_{i}^{k}-h_{ij}h^{jk}\neq \delta _{i}^{k}$. The reason is that (2.7) already
implies the existence of a basis of covariant and contravariant basic vector
fields $e_{i}$ and $\widetilde{e}^{j}$, such that $e_{i}\widetilde{e}%
^{j}=f_{i}^{j}$- in fact, this will be the essence of a proposition, which
shall soon be proved. Also, if $e^{j}$ is another system of basic fields for
which $e_{i}e^{j}=\delta _{i}^{j}$, then $\widetilde{e}^{j}=f_{k}^{j}e^{k}$
and the orthonormality condition can be written as (for $\alpha =\beta $) 
\begin{equation}
\delta _{i}^{j}=t_{i}^{\alpha }t_{\alpha }^{j}=\overline{t}_{i}^{\alpha
}e_{i}\widetilde{e}^{\alpha }\overline{t}_{\alpha }^{j}e_{\alpha }\widetilde{%
e}^{j}=\overline{t}_{i}^{\alpha }\overline{t}_{\alpha }^{j}f_{i}^{\alpha
}f_{\alpha }^{j}\text{ \ \ .}  \tag{2.16}
\end{equation}%
But the orthonormality condition is defined and should have one and the same
form in \textit{all reference frames}, including the reference frame $\left(
e_{\alpha },\widetilde{e}^{j}\right) $, in which the components of the
tetrad field are $\overline{t}_{i}^{\alpha }$. Consequently $f_{i}^{\alpha
}f_{\alpha }^{j}=1$, which however is in contradiction with the
arbitrariness in determining $f_{i}^{\alpha }$. The contradiction is due to
the assumption that the tensor transformation property (2.14) holds, and
since the expression in (2.14) equals $f_{i}^{\alpha }$, it \textit{should
not} transform as a tensor (note also that $f_{i}^{\alpha }\neq f_{\alpha
}^{i}$), at least for the investigated case of the projective transformation
(2.5). Also, the contradiction is that (2.15) is fulfilled for any $%
t_{i}^{\alpha }:=$\ $\frac{\partial x^{\alpha }}{\partial X^{i}}$ and $%
t_{\beta }^{j}:=$\ $\frac{\partial X^{j}}{\partial x^{\beta }}$, or in other
words - \textit{it should hold in any basis}, but at the same time \textit{%
we found a basis, in which (2.16) holds and not (2.15)}.

For the case of standard relativistic hydrodynamics, although $%
f_{i}^{k}=\delta _{i}^{k}-u_{i}u^{k}\neq \delta _{i}^{k}$, such a problem of
course will not appear because of the unit vector normalization $%
u^{i}u_{i}=1 $ for \textit{every vector field}, which is imposed \textit{%
apriori}.

Now it is easy to understand why and in \textit{what cases the distinction
between covariant and contravariant metric components will lead to an
inevitable introduction of two different connections }$\Gamma _{ij}^{k}$%
\textit{\ and }$P_{ij}^{k}$\textit{.} For the purpose, let us prove the
following statement:

\begin{proposition}
If $e_{1},e_{2},...,e_{n}$ is a basis of covariant vector fields and $%
f_{i}^{\alpha }$ are certain functions or constants, then a basis of
contravariant basic fields $\widetilde{e}^{\alpha _{1}},\widetilde{e}%
^{\alpha _{2}},...,\widetilde{e}^{\alpha _{n}}$ can be found so that for
each $i$ and $\alpha _{j}$ one has\ \ $e_{i}\widetilde{e}^{\alpha
_{j}}=f_{i}^{\alpha }$.
\end{proposition}

\bigskip This statement in fact is a generalization of the well-known
theorem from differential geometry that if a basis of (covariant) vector
fields is given then a dual basis of (contravariant) vector fields can be
found, so that the contravariant vector fields are the \textit{inverse ones
to the covariant} ones, i.e. $e_{i}\widetilde{e}^{\alpha _{j}}=\delta
_{i}^{\alpha }$.

The proof is very simple, but essentially based on the relation (2.13). If
the covariant basic vector fields are given, then the contravariant
connection components $\Gamma _{ij}^{k}$ will be known too. Since $%
f_{j,k}^{i}$ are derivatives of a function, one may take the expression
(2.13) \ \ \ \ \ \ \ \ $f_{j,k}^{i}=\Gamma _{jk}^{l}$ $%
f_{l}^{i}+P_{lk}^{i}f_{j}^{l}$ , which for the moment shall be treated as a
system of $n.[\frac{n(n+1)}{2}]$ \textit{linear algebraic equations with
respect to the (unknown) connection components }$P_{lk}^{i}$. A solution of
this system can be found for the connection components $P_{lk}^{i}$. Then
the condition for the parallel transport of the contravariant basic vector
fields $\nabla _{e_{\beta }}\widetilde{e}^{\alpha }=P_{\gamma \beta
}^{\alpha }$ $\widetilde{e}^{\gamma }$ can be written as $\partial _{\beta }%
\widetilde{e}^{\alpha }=P_{\gamma \beta }^{\alpha }$ $\widetilde{e}^{\gamma
} $ and considered as a system of $n$\textit{\ ordinary differential
equations with respect to the components }$\widetilde{e}^{\alpha }$. From
this system, the unique solution for $\widetilde{e}^{\alpha _{1}},\widetilde{%
e}^{\alpha _{2}},...,\widetilde{e}^{\alpha _{n}}$ can be found up to
integration constants, obtained after the integration of the differential
equations.

After proving this proposition, the difference between standard relativistic
hydrodynamics and \textit{"modified" relativistic hydrodynamics with a
variable length} can be easily understood. In the first case, the right-hand
side in $p_{ij}p^{jk}=\delta _{i}^{k}-u_{i}u^{k}=f_{i}^{k}\neq \delta
_{i}^{k}$ transforms as a tensor, which is ensured also by normalization
property $u_{i}u^{i}=1$.Therefore (2.13) and the proposition will not hold,
so the contravariant basic vector fields are determined in the standard way $%
e_{i}e^{j}=\delta _{i}^{j}$ and more importantly, they \textit{cannot be
determined} in another way, in spite of the fact that again $f_{i}^{k}\neq
\delta _{i}^{k}$.

In the second case, the situation is just the opposite - the right-hand side
of $p_{ij}p^{jk}=\delta _{i}^{k}-\frac{1}{e}u_{i}u^{k}=f_{i}^{k}\neq \delta
_{i}^{k}$ transforms not as a tensor because of the "normalization" factor $%
\frac{1}{e}$, the proposition holds and thus the basic vector fields are
determined as $e_{i}\widetilde{e}^{j}=f_{i}^{j}$. \textit{Therefore, the
treatment of relativistic hydrodynamics with "variable length" should be
within the GTCCMC}.

In the present case, the introduced new connection (2.4) \textit{should not
be identified} with the connection $P_{\alpha \beta }^{\gamma }$, since the
connection $\widetilde{\Gamma }_{kl}^{s}\equiv \widetilde{g}^{is}\Gamma
_{i;kl}$ \ is \ introduced by means of modifying the contravariant tensor
and not on the base of any separate parallel transport. Moreover, the
connection $\widetilde{\Gamma }_{kl}^{s}$ turns out to be a linear
combination of the Christoffell connection components \ $\Gamma _{\alpha
\beta }^{\gamma }$, and the relation between them is not of the type (2.13).
In such a way, there will not be a contradiction with the case when the%
\textit{\ two connections }$\Gamma _{\alpha \beta }^{\gamma }$\textit{\ and }%
$\widetilde{\Gamma }_{kl}^{s}$ \textit{are not defined as separate ones},
since later on, in deriving the cubic algebraic equation in the general case
and for the case $\widetilde{g}^{jk}=dX^{j}dX^{k}$ also, it would be
supposed that $\widetilde{g}^{is}$ is a tensor. This would mean (from $%
\widetilde{g}^{is}g_{im}\equiv f_{m}^{s}(\mathbf{x})$) that $f_{m}^{s}(%
\mathbf{x})$ will also be a (mixed) tensor quantity, and therefore the
covariant differentiation of $e_{\alpha }e^{\beta }\equiv f_{\alpha }^{\beta
}(x)$ will not produce any new relation.

\section{\protect\bigskip\ BASIC \ ALGEBRAIC \ EQUATIONS \ IN \ GRAVITY \
THEORY. TENSOR\ \ LENGTH\ SCALE}

Now if one applies again the new definition $\widetilde{g}^{ij}\equiv
dX^{i}dX^{j}$ of the contravariant tensor with respect to the Ricci tensor,
then the following fourth - degree algebraic equation can be obtained 
\begin{equation*}
R_{ik}=dX^{l}\left[ g_{is,l}\frac{\partial (dX^{s})}{\partial x^{k}}-\frac{1%
}{2}pg_{ik,l}+\frac{1}{2}g_{il,s}\frac{\partial (dX^{s})}{\partial x^{k}}%
\right] +
\end{equation*}%
\begin{equation}
+\frac{1}{2}dX^{l}dX^{m}dX^{r}dX^{s}\left[
g_{m[k,t}g_{l]r,i}+g_{i[l,t}g_{mr,k]}+2g_{t[k,i}g_{mr,l]}\right] \text{ \ \
\ \ \ ,}  \tag{3.1}
\end{equation}%
where $p$ is the scalar quantity

\begin{equation}
p\equiv div(dX)\equiv \frac{\partial (dX^{l})}{\partial x^{l}}\text{,} 
\tag{3.2}
\end{equation}%
which \textquotedblright measures\textquotedblright\ the divergency of the
vector field $dX$. The algebraic variety of the equation consists of the
differentials $dX^{i\text{ }}$ and their derivatives $\frac{\partial (dX^{s})%
}{\partial x^{k}}$.

In the same spirit, one can investigate the problem whether the
gravitational Lagrangian in terms of the new contravariant tensor can be
equal to the standard representation of the gravitational Lagrangian. This
standard \textit{(first) representation} of the gravitational Lagrangian is
based on the standard Christoffell connection $\Gamma _{ij}^{k}$ (given by
formulae (2.1)), the Ricci tensor $R_{ik}$ (formulae (2.2)) and the \textit{%
other contravariant tensor} $\widetilde{g}^{ij}=dX^{i}dX^{j}$ 
\begin{equation}
L_{1}=-\sqrt{-g}\widetilde{g}^{ik}R_{ik}=-\sqrt{-g}dX^{i}dX^{k}R_{ik}\text{
\ .}  \tag{3.3}
\end{equation}%
In the \textit{second representation }the\textit{\ }Christoffell connection $%
\widetilde{\Gamma }_{ij}^{k}$ and the Ricci tensor $\widetilde{R}_{ik}$%
\textit{\ }are "tilda" quantities, meaning that the "tilda" Christoffell
connection is determined by formulae (2.4) with the new contravariant tensor 
$\widetilde{g}^{ij}=dX^{i}dX^{j}$ and the "tilda" Ricci tensor $\widetilde{R}%
_{ik}$ - by formulae (2.2), but with the "tilda" connection $\widetilde{%
\Gamma }_{ij}^{k}$ instead of the usual Christoffell connection $\Gamma
_{ij}^{k}$. Thus the expression for the \textit{second representation} of
the gravitational Lagrangian acquires the form 
\begin{equation}
L_{2}=-\sqrt{-g}\widetilde{g}^{il}\widetilde{R}_{il}=-\sqrt{-g}%
dX^{i}dX^{l}\{p\Gamma _{il}^{r}g_{kr}dX^{k}-\Gamma
_{ik}^{r}g_{lr}d^{2}X^{k}-\Gamma _{l(i}^{r}g_{k)r}d^{2}X^{k}\}\text{ .} 
\tag{3.4}
\end{equation}%
The condition for the \textit{equivalence of the two representations} (i.e. $%
L_{1}=L_{2}$) gives a \textit{cubic algebraic equation} with respect to the
algebraic variety of the first differential $dX^{i}$ and the second one $%
d^{2}X^{i}$ [10] 
\begin{equation}
dX^{i}dX^{l}\left( p\Gamma _{il}^{r}g_{kr}dX^{k}-\Gamma
_{ik}^{r}g_{lr}d^{2}X^{k}-\Gamma _{l(i}^{r}g_{k)r}d^{2}X^{k}\right)
-dX^{i}dX^{l}R_{il}=0\text{ \ \ \ \ .}  \tag{3.5}
\end{equation}%
Note the following essential peculiarity of the second representation (3.4)
- due to the choice of the "modified" contravariant tensor, the second
quadratic term with the "tilda" connection in the expression for $\widetilde{%
R}_{ij}$ is equal to zero%
\begin{equation*}
-\sqrt{-g}dX^{i}dX^{k}(\widetilde{\Gamma }_{ik}^{l}\widetilde{\Gamma }%
_{lm}^{m}-\widetilde{\Gamma }_{il}^{m}\widetilde{\Gamma }_{km}^{l})\text{ }=-%
\frac{1}{2}\sqrt{-g}%
dX^{i}dX^{k}dX^{l}dX^{m}(-dg_{lm}dX^{s}g_{ks,i}-dg_{ik}dX^{r}g_{mr,l}+
\end{equation*}%
\begin{equation*}
+dg_{il}dX^{r}g_{mr,k}+dg_{km}dX^{s}g_{ls,i})-
\end{equation*}%
\begin{equation}
-\sqrt{-g}%
dX^{i}dX^{k}dX^{l}dX^{m}dX^{s}dX^{r}(g_{ks,i}g_{mr,l}-g_{ls,i}g_{mr,k})=0%
\text{.}  \tag{3.6}
\end{equation}%
The $\mathit{first\ differential}$ $dg_{ij}$ in (3.6) is represented as $%
dg_{ij}\equiv \frac{\partial g_{ij}}{\partial x^{s}}dX^{s}\equiv \Gamma
_{s(i}^{r}g_{j)r}dX^{s}$.

\textbf{\ }Following the same approach, in [10] the Einstein's equations in
vacuum for the general case were derived under the assumption that the
contravariant metric tensor components are the "tilda" ones: 
\begin{equation*}
0=\widetilde{R}_{ij}-\frac{1}{2}g_{ij}\widetilde{R}=\widetilde{R}_{ij}-\frac{%
1}{2}g_{ij}dX^{m}dX^{n}\widetilde{R}_{mn}=
\end{equation*}%
\begin{equation*}
=-\frac{1}{2}pg_{ij}\Gamma _{mn}^{r}g_{kr}dX^{k}dX^{m}dX^{n}+\frac{1}{2}%
g_{ij}(\Gamma _{km}^{r}g_{nr}+\Gamma
_{n(m}^{r}g_{k)r})d^{2}X^{k}dX^{m}dX^{n}+
\end{equation*}%
\begin{equation}
+p\Gamma _{ij}^{r}g_{kr}dX^{k}-(\Gamma _{ik}^{r}g_{jr}+\Gamma
_{j(i}^{r}g_{k)r})d^{2}X^{k}\text{ \ \ \ .}  \tag{3.7 }
\end{equation}

\bigskip This equation represents again a system of cubic equations. In
addition, if the differentials $dX^{i}$\ and $d^{2}X^{i}$\ are known, but
not the covariant tensor $g_{ij}$, the same equation can be considered also
as a \ cubic algebraic equation with respect to the algebraic variety of the
metric tensor components $g_{ij}$\ and their first derivatives $g_{ij,k}$.

It might be thought that the definite choice of the contravariant tensor is
a serious restriction, in view of the fact that the second derivatives of
the covariant tensor components $g_{ij,kl}$ are not present in the equation.
This is indeed so, because the algebraic structure of the equation is
simpler to deal with in comparison with the general case, and so it is
easier to implement the algorithm for parametrization, developed in [10].
But there is one argument in favour of this choice (although the case for an
arbitrary contravariant tensor is also more important) - since the metric
can be expressed as $ds^{2}=l(x)=g_{ij}dX^{i}dX^{j}$ (consequently $%
dX^{i}dX^{j}=l(x)g^{ij}$), the obtained cubic algebraic equations (3.5) and
(3.7) can be considered with regard also to the length function $l(x)$.
Since for Einsteinian gravity $g_{ij}g^{jk}=\delta _{i}^{k}$ (i.e. $g^{jk}=%
\widetilde{g}^{jk}=dX^{j}dX^{k}$), then for this case the length function is
"postulated" to be $l=1$. Note that this choice of the contravariant tensor $%
\widetilde{g}^{jk}$ in the form of a factorized product is a partial (and
not a general) choice, but further it shall be shown that the cubic equation
for reparametrization  invariance of the gravitational Lagrangian can be
written also for a generally chosen tensor $\widetilde{g}^{jk}$. Then from $%
g_{ij}\widetilde{g}^{jk}=l_{i}^{k}$ and $l_{i}^{k}=l\delta _{i}^{k}$, the
length function is again recovered. The important point here is that the
length function can also be obtained as a solution of the cubic equation,
and thus in more general theories of gravity solutions with $l\neq 1$ may
exit. In fact, for a general contravariant tensor $\widetilde{g}%
^{ij}=l_{k}^{i}g^{kj}\neq dX^{i}dX^{j}$, it would be natural to propose to
call $l_{k}^{i}$ a \textit{"tensor length scale", }and the previously
defined length function $l(x)$ is a partial case of the tensor length scale
for $l_{j}^{i}=l\delta _{j}^{i}$. The \textit{physical meaning} of the
notion of tensor length scale is simple - in the different directions (i.e.
for different $i$ and $j$) the length scale is \ different. In particular,
some motivation for this comes from Witten's paper [46], where in discussing
some aspects of weakly coupled heterotic string theory (when there is just
one string couplings ) and the obtained too large bound on the Newton's
constant it was remarked that \textit{\textquotedblright the problem might
be ameliorated by considering an anisotropic Calabi - Yau with a scale }$%
\sqrt{\alpha ^{^{\prime }}}$\textit{\ in }$d$\textit{\ directions and }$%
\frac{1}{M_{GUT}}$\textit{\ in }$(6-d)$\textit{\
directions\textquotedblright }. For example, this can be realized if one
takes 
\begin{equation}
l_{i}^{k}=g_{ij}dX^{j}dX^{k}=l_{1}\delta _{i}^{k}\text{ \ for \ }%
i,j,k=1,....,d\text{ \ \ \ \ ,}  \tag{3.8}
\end{equation}%
\begin{equation}
l_{a}^{b}=g_{ac}dX^{c}dX^{b}=l_{2}\delta _{a}^{b}\text{ \ for \ }%
a,b,c=d+1,....,6\text{\ \ \ \ .}  \tag{3.9}
\end{equation}%
Note also the justification for the name "tensor length scale" - if \ $%
l_{k}^{i}$ is a tensor quantity, so will be the "modified" contravariant
tensor $\widetilde{g}^{ij}=l_{k}^{i}g^{kj}$, and consequently in accord with
section 2 there will be no need for the introduction of a new covariant
connection $P_{ij}^{k}$. And this is indeed the case, because the relation
between the two connections $\Gamma _{ij}^{k}$ and $\widetilde{\Gamma }%
_{ij}^{k}$ is given by formulae (2.4) $\widetilde{\Gamma }_{kl}^{s}:=%
\widetilde{g}^{is}g_{im}\Gamma _{kl}^{m}$. In other words, since these two
connections are not considered to be "separately introduced" and so they do
not depend on one another by means of the equality (2.13), this particular
investigated case \textit{does not fall} within the classification of spaces
with covariant and contravariant metrics and connections (Table I in a
previous paper [47]). This is an important "terminological" clarification,
since it turns out that it is possible to have a theory with \textit{%
(separate) covariant and contravariant metrics, but not (with separate)
connections as well. And such a theory is fully consistent from a
mathematical point of view, as demonstrated above. }However, at this point
it is important to clarify what is meant by "a theory with \textit{%
(separate) covariant and contravariant metrics}" - it should be understood 
\textit{only} with respect to the metrics $g_{\mu \nu }$ and \textit{\ }$%
\widetilde{g}^{is}$. But if we take the contravariant metric $\widetilde{g}%
^{is}$ (and ignore for the moment the metric $g_{\mu \nu }$), then from the
equality $\overline{g}_{ij}\widetilde{g}^{jk}=\delta _{i}^{k}$ one can
determine an inverse to the contravariant metric $\widetilde{g}^{is}$ 
\textit{new covariant metric} $\overline{g}_{ij}$, and consequently, the
following \textit{new contravariant connection} $\overline{\Gamma }_{kl}^{s}$
can also be determined 
\begin{equation}
\overline{\Gamma }_{kl}^{s}:=\widetilde{g}^{is}\overline{\Gamma }_{i;kl}=%
\widetilde{g}^{is}\overline{g}_{im}\overline{\Gamma }_{kl}^{m}=\frac{1}{2}%
\widetilde{g}^{is}(\overline{g}_{ik,l}+\overline{g}_{il,k}-\overline{g}%
_{kl,i})\text{.}  \tag{3.10}
\end{equation}%
Evidently, with respect to the metric $\overline{g}_{ij}$ (and its inverse
contravariant one $\widetilde{g}^{jk}$),we have the usual gravitational
theory with the contravariant $\overline{\Gamma }_{\alpha \beta }^{\gamma }$
and covariant $\overline{P}_{\alpha \beta }^{\gamma }$ connections 
\begin{equation}
\nabla _{e_{\beta }}e_{\alpha }=\overline{\Gamma }_{\alpha \beta }^{\gamma
}e_{\gamma }\text{ ; \ }\nabla _{e_{\beta }}e^{\alpha }=\overline{P}_{\gamma
\beta }^{\alpha }e^{\gamma }\text{ ; \ }\overline{P}_{\gamma \beta }^{\alpha
}=-\overline{\Gamma }_{\alpha \beta }^{\gamma }\text{ \ \ .}  \tag{3.11}
\end{equation}%
However, although with respect to the metrics $g_{\mu \nu }$ and \textit{\ }$%
\widetilde{g}^{is}$ and the connections $\Gamma _{ij}^{k}$ and $\widetilde{%
\Gamma }_{kl}^{s}:=\widetilde{g}^{is}g_{im}\Gamma _{kl}^{m}$ (3.4)\ the
theory is with \textit{\ covariant and contravariant metrics (only), }%
connections $\overline{\Gamma }_{ij}^{k}$ and $\widetilde{\Gamma }_{kl}^{s}$
can be determined (by means of the additional metric $\overline{g}_{\mu \nu }
$) in the following way 
\begin{equation}
\nabla _{e_{\beta }}e_{\alpha }=\overline{\Gamma }_{\alpha \beta }^{\gamma
}e_{\gamma }\text{ ; }\ \ \ \ \ \ \ \ \text{\ }\nabla _{e_{\beta }}e^{\alpha
}=\widetilde{\Gamma }_{\gamma \beta }^{\alpha }e^{\gamma }\text{ , \  \ \ } 
\tag{3.12}
\end{equation}%
so that with respect to these connections the theory can be considered a
GTCCMC. \textit{This also means that Table I in [47] correctly does not
account for theories with different covariant and contravariant metrics
only, because the different GTCCMC are in principle} \textit{with different
covariant and contravariant metrics and with different connections.} 

The purpose of the present paper further will be:\ \textit{how can one
extend the proposed in [10] approach for the definitely determined
contravariant metric components to the case of a generally defined
contravariant tensor }$\widetilde{g}^{ij}\neq dX^{i}dX^{j}$\textit{? }

\section{ INTERSECTING ALGEBRAIC VARIETIES AND STANDARD (EINSTEINIAN)
GRAVITY THEORY}

A more general theory with the definition of the contravariant tensor as $%
\widetilde{g}^{ij}\equiv dX^{i}dX^{j}$ should \ contain in itself the
standard gravitational theory with $g_{ij}g^{jk}=\delta _{i}^{k}$. From a
mathematical point of view, this should be performed by considering the
intersection [19] of the cubic algebraic equations (3.7)\ \ with the system
of \ $n^{2}$ quadratic algebraic equations for the algebraic variety of the $%
n$ variables 
\begin{equation}
g_{ij}dX^{j}dX^{k}=\delta _{i}^{k}\text{ \ \ .}  \tag{4.1}
\end{equation}%
In its general form $g_{ij}\widetilde{g}^{jk}=\delta _{i}^{k}$ with an
arbitrary contravariant tensor $\widetilde{g}^{jk}$, this system \ can also
be considered together with the Einstein's "algebraic" system of equations,
which in the next section shall be derived for a \textit{generally defined
contravariant tensor.} From an algebro - geometric point of view, this is
the problem about the intersection of the Einstein's algebraic equations
with the system of $n^{2}$ \ (linear) hypersurfaces for the $\left[ \left( 
\begin{array}{c}
n \\ 
2%
\end{array}%
\right) +n\right] $\ contravariant variables, if the covariant tensor
components are given. Since the derived Einstein's algebraic equations are
again cubic ones with respect to the contravatiant metric components, this
is an analogue to the well - known problem in algebraic geometry about the
intersection of a (two-dimensional)\ cubic curve with a straight line.
However, in the present case the straight line and the cubic curve are 
\textit{multi - dimensional ones}, which is a substantial difference from
the standard case.

The standardly known solutions of the Einstein's equations can be obtained
as an intersection variety of the Einstein's algebraic equations with the
system $g_{ij}\widetilde{g}^{jk}=\delta _{i}^{k}$ . However, the strict
mathematical proof that such an intersection will give the known solutions
is still lacking.

It might seem that the system of equations (4.1) does not have solutions
with respect to $g_{ij\text{ }}$(and thus no solutions of the Einstein's
equations can be found for the standard case), since the determinant $%
det\parallel dX^{i}dX^{j}\parallel _{i,j=1,..n}=0$ equals to zero! In
another paper it will be proved that such a matrix operator system of
equations [20] $Y_{ij}g^{jk}=\delta _{i}^{k}$ with unknown variables $%
Y_{ij}\equiv g_{ij}$ (which is not a system of linear algebraic equations,
but instead a system of \textit{matrix equations}) can be transformed to a
system of linear algebraic equations $\widetilde{A}_{ij}\widetilde{Y}^{j}$ $%
=T_{i}$ ($T_{i}$ is a \textbf{\ }vector - column). This system always has a
solution \textit{at least for some} of the variables - the others may be
determined arbitrarily. Therefore, solutions will exist and will be
well-determined even in the case of a zero determinant.\textbf{\ }

\section{ ALGEBRAIC \ EQUATIONS \ FOR \ A \ GENERAL \ CONTRAVARIANT \ METRIC
\ TENSOR}

Let us write down the algebraic equations for all admissable
parametrizations of the gravitational Lagrangian for the generally defined
contravariant tensor $\widetilde{g}^{ij}$, following \textit{the same
prescription} as in section 3, where the equality of the two representations
of the gravitational Lagrangian has been supposed: 
\begin{equation*}
\widetilde{g}^{i[k}\widetilde{g}_{,l}^{l]s}\Gamma _{ik}^{r}g_{rs}+\widetilde{%
g}^{i[k}\widetilde{g}^{l]s}\left( \Gamma _{ik}^{r}g_{rs}\right) _{,l}+
\end{equation*}%
\begin{equation}
+\widetilde{g}^{ik}\widetilde{g}^{ls}\widetilde{g}^{mr}g_{pr}g_{qs}\left(
\Gamma _{ik}^{q}\Gamma _{lm}^{p}-\Gamma _{il}^{p}\Gamma _{km}^{q}\right) -R=0%
\text{ \ \ \ \ .}  \tag{5.1}
\end{equation}%
This equation is again a \textit{cubic algebraic equation} with \ respect to
the algebraic variety of the variables $\widetilde{g}^{ij}$ and $\widetilde{g%
}_{,k}^{ij}$, and the number of variables in the present case is much
greater than in the previous case for the contravariant tensor $\widetilde{g}%
^{ij}\equiv dX^{i}dX^{j}$ . At  the same time, this equation is a \textit{%
fourth - degree} algebraic equation with respect to the covariant metric
tensor $g_{ij}$ and its first and second partial derivatives. With respect
to the algebraic variety of all the variables $\widetilde{g}^{ij}$, $%
\widetilde{g}_{,k}^{ij}$, $g_{ij}$, $g_{ij,k}$, $g_{ij,kl}$, the above
algebraic equation is of \textit{seventh order} and with coefficient
functions, due to the presence of the terms with the affine connection $%
\Gamma _{ik}^{q}$ and its derivatives, which contain the contravariant
tensor $g^{ij}$ and $g_{,k}^{ij}$.

If the connection is assumed to be the \textquotedblright tilda
connection\textquotedblright\ $\widetilde{\Gamma }_{kl}^{s}\equiv
dX^{i}dX^{s}\Gamma _{i;kl}$, then the same equation can be regarded as a 
\textit{sixth - degree} equation with respect to the algebraic variety of $%
dX^{i}$ and its derivatives.

Similarly, the Einstein's equations can be written as a system of \ \textit{%
third - degree algebraic equations} with respect to the (generally chosen)
contravariant variables and their derivatives 
\begin{equation*}
0=\widetilde{R}_{ij}-\frac{1}{2}g_{ij}\widetilde{R}=
\end{equation*}%
\begin{equation*}
=\widetilde{g}^{lr}(\Gamma _{r;i[j}),_{l]}+\widetilde{g}_{,[l}^{lr}\Gamma
_{r;ij]}+\widetilde{g}^{lr}\widetilde{g}^{ms}(\Gamma _{r;ij}\Gamma
_{s;lm}-\Gamma _{s;il}\Gamma _{r;km})-
\end{equation*}%
\begin{equation*}
-\frac{1}{2}g_{ij}\widetilde{g}^{m[k}\widetilde{g}_{,l}^{l]s}\Gamma
_{mk}^{r}g_{rs}-\frac{1}{2}g_{ij}\widetilde{g}^{m[k}\widetilde{g}%
^{l]s}\left( \Gamma _{mk}^{r}g_{rs}\right) _{,l}-
\end{equation*}%
\begin{equation}
-\frac{1}{2}g_{ij}\widetilde{g}^{nk}\widetilde{g}^{ls}\widetilde{g}%
^{mr}g_{pr}g_{qs}\left( \Gamma _{nk}^{q}\Gamma _{lm}^{p}-\Gamma
_{nl}^{p}\Gamma _{km}^{q}\right) \text{ \ .}  \tag{5.2}
\end{equation}%
Interestingly, the same system of equations can be considered as a system of
\ \textit{fifth - degree} equations with respect to the covariant variables
(which is the difference from the previous case). The mathematical treatment
of fifth - degree equations is known since the time of Felix Klein's famous
monograph [25], published in 1884. A way for resolution of such equations on
the base of earlier developed approaches by means of reducing the fifth -
degree equations to the so called modular equation has been presented in the
more recent \ monograph of \ Prasolov and Solov'yev [9]. Some other methods
for solution of third-, fifth- and higher- order \ algebraic equations have
been given in [26, 27]. A complete description of elliptic, theta and
modular functions has been given in the old monographs [28, 29]. Also,
solutions of $n-$ th degree algebraic equations in theta - constants [30]
and in special functions [31] are interesting in view of the not yet proven
hypothesis in the paper by Kraniotis and Whitehouse [8] that \textit{%
\textquotedblright all nonlinear solutions of general relativity are
expresed in terms of theta - functions, associated with Riemann -
surfaces\textquotedblright }. 

The basic knowledge about the parametrization of a cubic algebraic equation
with the Weierstrass function and its derivative, which shall be extensively
used in the subsequent parts of this paper,   is given in almost all basic
textbooks on elliptic functions [9, 11, 32, 33] and many others. However,
the most complete, detailed and exhaustive knowledge about elliptic
functions and automorphic forms is contained in the two two - volume \ books
[34, 35] of \ Felix Klein and Robert Fricke, written more than 100 years
ago. More specific and advanced topics on elliptic curves from a
mathematical point of view such as the group of rational points, cubic
curves over finite fields, families of elliptic curves and torsion points
and etc. are contained in the monographs [36, 37]. A very understandable
exposition of the classical topics on cubic algebraic curves and at the same
time the most contemporary issues such as the Mordell's and Dirichlet's
theorems and $L$ functions, modular forms and theories of Eichler - Shimura
are given in the book of Knapp [38], which can be used for first acquintance
in these topics. A consistent, modern and full exposition of elliptic curves
in the language of modern mathematics is given in the (two consequent)\
monographs of Silverman [39, 40]. A classical and very understandable
exposition of the relation of elliptic curves with modular forms is given in
[41], also in [42]. From a modern standpoint the relation of elliptic curves
with number theory and modular forms is given in the review articles of
Cohen and Don Zagier in [43], also introductory knowledge on hyperelliptic
integrals, compact Riemann surfaces and Abelian varieties are presented in
the review article by Bost also in [43].

Two other important problems can be pointed out with reference to algebraic
equations in gravity theory:

1. One can find solutions of the system of Einstein's equations not as
solutions of a system of nonlinear differential equations, but as elements
of an algebraic variety, satisfying the Einstein's algebraic equations. The
important new moment is that this gives an opportunity to find solutions of
\ the Einstein's equations both for the components of the covariant metric
tensor $g_{ij}$\ and for the contravariant ones $\widetilde{g}^{jk}$. This
means that solutions may exist for which $g_{ij}\widetilde{g}^{jk}\neq
\delta _{i}^{k}$. In other words, a classification of the solutions of the
Einstein's equations can be performed in an entirely new and nontrivial
manner - under a given contravariant tensor, the covariant tensor and its
derivatives have to be found from the algebraic equation, or under a given
covariant tensor, the contravariant tensor and its derivatives can be found.

2. The condition for the zero - covariant derivative of the covariant metric
tensor $\nabla _{k}g_{ij}=0$ and of the contravariant metric tensor $\nabla
_{k}\widetilde{g}^{ij}=0$ can be written in the form of the following cubic
algebraic equations\textbf{\ }with respect to the variables $g_{ij}$, $%
g_{ij,k}$ and $\widetilde{g}^{ls}$ \textbf{:} 
\begin{equation}
\nabla _{k}g_{ij}\equiv g_{ij,k}-\widetilde{\Gamma }%
_{k(i}^{l}g_{j)l}=g_{ij,k}-\widetilde{g}^{ls}\Gamma _{s;k(i}g_{j)l}=0 
\tag{5.3}
\end{equation}%
and 
\begin{equation}
0=\nabla _{k}\widetilde{g}^{ij}=\widetilde{g}_{,k}^{ij}+\widetilde{g}^{r(i}%
\widetilde{g}^{j)s}\Gamma _{r;sk}\text{ \ \ \ \ .}  \tag{5.4}
\end{equation}%
The first equation (5.3) is linear with respect to $\widetilde{g}^{ls}$ and
quadratic with respect to $g_{ij}$, $g_{ij,k}$, while the second equation
(5.3) is linear with respect to $g_{ij}$, $g_{ij,k}$ and quadratic with
respect to $\widetilde{g}^{ls}$.

Since the treatment of the above cubic algebraic equations is based on
singling out one variable, let us rewrite equation (5.1) for the effective
parametrization of the gravitational action for the case of diagonal metrics 
$g_{\beta \beta }$ and $\widetilde{g}^{\alpha \alpha }$, singling out the
variable $\widetilde{g}^{44}$: 
\begin{equation*}
A(\widetilde{g}^{44})^{3}+B_{\alpha }(\widetilde{g}^{44})^{2}\widetilde{g}%
^{\alpha \alpha }+C_{\alpha \alpha }\widetilde{g}^{44}\widetilde{g}^{\alpha
\alpha }+(\Gamma _{44}^{\alpha }g_{\alpha \alpha })\widetilde{g}^{44}%
\widetilde{g}_{,\alpha }^{\alpha \alpha }+
\end{equation*}%
\begin{equation}
+D_{\alpha \gamma }\widetilde{g}^{44}\widetilde{g}^{\alpha \alpha }%
\widetilde{g}^{\gamma \gamma }+F_{\alpha \gamma }=0\text{ \ \ \ \ \ ,} 
\tag{5.5}
\end{equation}%
where the coefficient functions $A$, $B_{\alpha }$, $C_{\alpha \alpha }$, $%
D_{\alpha \gamma }$ and the free term $F_{\alpha \gamma }$ denotes an
expression, depending on the covariant metric tensor and the affine
connection. In (5.5) the Greek indices run the values $\alpha ,\beta ,\gamma
=1,2,3$, while all the other indices run from $1$ to $4$.

The representation (5.5) of the cubic equation is the starting point for the
parametrization with the Weierstrass function, which will be performed
elsewhere, following the algorithm in the paper [10]. In the second part of
this paper, this would be performed for the case a \textit{multivariable
cubic algebraic equation }(although again within the framework of the
factorizing approximation $\widetilde{g}^{ij}\equiv dX^{i}dX^{j}$) and this
is \textit{entirely different} from the standardly known case in algebraic
geometry of parametrization of a \textit{two-dimensional cubic algebraic
equation} in its parametrizable form.

\section{\protect\bigskip DISCUSSION}

\bigskip In this paper we continued the investigation of cubic algebraic
equations in gravity theory, which has been initiated in a previous paper
[10].

Unlike in the paper [10], where the treatment of cubic algebraic equations
has been restricted only to the choice of the contravariant tensor $%
\widetilde{g}^{ij}=dX^{i}dX^{j}$, in the present paper it has been
demonstrated that under a more general choice of $\widetilde{g}^{ij}$, there
is a wide variety of algebraic equations of various order, among which an
important role play the cubic equations. Their derivation is based on two
important initial assumptions:

1. The covariant and contravariant metric components are treated
independently, which is a natural approach in the framework of \textit{%
affine geometry} [15 - 18].

2. Under the above assumption, the gravitational Lagrangian (or Ricci
tensor) should remain the same as in the standard gravitational theory with
inverse contravariant metric tensor components.

It will be proved in Appendix A that the new connection $\widetilde{\Gamma }%
_{ij}^{k}=\frac{1}{2}dX^{k}dX^{s}(g_{js,i}+g_{is,j}-g_{ij,s})$ has again an
affine connection transformation property, provided that a complicated
system of nonlinear differential equations are satisfied. This system is
expected to have a broad class of solutions.

The proposed approach allows to treat the Einstein's equations as algebraic
equations, and thus to search for separate classes of solutions for the
covariant and contravariant metric tensor components\textbf{.} It can be
supposed also that the existence of such separate classes of solutions might
have some interesting and unexplored until now physical consequences. Some
of them will be demonstrated in reference to theories with extra dimensions,
but no doubt the physical applications are much more numerous.

Also, it has been shown that the "transition" to the standard Einsteinian
theory \ of gravity can be performed by investigating the intersection with
the corresponding algebraic equations.

\subsection*{\protect\bigskip Acknowledgments}

The author is grateful to Dr. I. Pestov (BLTP, JINR, Russia), Dr. D. M.
Mladenov (Theor.Physics Departm., Fac. of Physics, Sofia Univ., Bulgaria),
and especially to Prof. V. V. Nesterenko (BLTP, JINR, Russia), Dr. O.
Santillan (IAFE, Buenos Aires, Argentina) and Prof. Sawa Manoff (INRNE, BAS,
Bulgaria) for valuable comments, discussions and critical remarks. \ 

This paper is written in memory of \ Prof. S. S. Manoff (1943 - 27.05.2005)
- a specialist in classical gravitational theory. \ 

The author is grateful also to Dr. G. V. Kraniotis (Max Planck Inst.,
Munich, Germany) for sending to me his published paper (ref. [8]).

\section{\protect\bigskip APPENDIX\ \ A: SOME \ PROPERTIES \ OF \ THE \
NEWLY \ INTRODUCED \ CONNECTION $\widetilde{\Gamma }_{ij}^{k}=\frac{1}{2}%
dX^{k}dX^{l}(g_{jl,i}+g_{il,j}-g_{ij,l})$\ }

\subsection*{\protect\bigskip A1: \ A \ PROOF \ OF \ THE \ AFFINE \
TRANSFORMATION \ LAW \ FOR \ THE \ CONNECTION \ $\widetilde{\Gamma }%
_{ij}^{k} $}

\bigskip

\subsubsection*{\protect\bigskip A1.1 \ THE\ \ NECESSARY \ CONDITION\ }

Next we proceed with the proof that the defined ( in the preceeding paper
[10] also) connection 
\begin{equation}
\widetilde{\Gamma }_{ij}^{k}=\frac{1}{2}%
dX^{k}dX^{l}(g_{jl,i}+g_{il,j}-g_{ij,l})=dX^{k}dX^{r}g_{sr}(X)\Gamma
_{ij}^{s}(X)  \tag{A1}
\end{equation}%
has the transformation property of an affine connection under the coordinate
transformations $X^{i}=X^{i}(x^{1},x^{2},...,x^{n})$.

From the defining equation (A1) for $\widetilde{\Gamma }_{ij}^{k}$, the
tensor transformation property for $g_{ij}^{^{\prime }}(\mathbf{X})$ 
\begin{equation}
g_{ij}^{^{\prime }}(\mathbf{X})=\frac{\partial x^{k}}{\partial X^{i}}\frac{%
\partial x^{l}}{\partial X^{j}}g_{kl}(\mathbf{x})\text{ \ \ ,}  \tag{A2}
\end{equation}%
the affine transformation law for the "usual" connection $\Gamma _{ij}^{k}$ 
\begin{equation}
\Gamma _{ij}^{k^{\prime }}(\mathbf{X})=\Gamma _{np}^{m}(\mathbf{x})\frac{%
\partial X^{k}}{\partial x^{m}}\frac{\partial x^{n}}{\partial X^{i}}\frac{%
\partial x^{p}}{\partial X^{j}}+\frac{\partial ^{2}x^{m}}{\partial
X^{i}\partial X^{j}}\frac{\partial X^{k}}{\partial x^{m}}  \tag{A3}
\end{equation}%
and from the expressions for the differentials $dX^{k}$ and $dX^{r}$ we may
write down 
\begin{equation}
\widetilde{\Gamma }_{ij}^{k^{\prime }}(\mathbf{X})==dX^{k}(\mathbf{X})dX^{r}(%
\mathbf{X})g_{sr}^{^{\prime }}(\mathbf{X})\Gamma _{ij}^{s^{\prime }}(\mathbf{%
X})=  \tag{A4}
\end{equation}%
\begin{equation}
=\Gamma _{np}^{m}(\mathbf{x})\frac{\partial X^{k}}{\partial x^{\alpha }}%
\frac{\partial x^{n}}{\partial X^{i}}\frac{\partial x^{p}}{\partial X^{j}}%
g_{m\beta }(\mathbf{x})dx^{\alpha }dx^{\beta }+\frac{\partial ^{2}x^{m}}{%
\partial X^{i}\partial X^{j}}\frac{\partial X^{k}}{\partial x^{\alpha }}%
g_{m\beta }(\mathbf{x})dx^{\alpha }dx^{\beta }\text{ .}  \tag{A5}
\end{equation}

On the other hand, if $\widetilde{\Gamma }_{ij}^{k}(\mathbf{X})$ is an
affine connection, then it should satisfy the affine connection
transformation law (A3) 
\begin{equation}
\widetilde{\Gamma }_{ij}^{k^{\prime }}(\mathbf{X})=\widetilde{\Gamma }%
_{np}^{m}(\mathbf{x})\frac{\partial X^{k}}{\partial x^{m}}\frac{\partial
x^{n}}{\partial X^{i}}\frac{\partial x^{p}}{\partial X^{j}}+\frac{\partial
^{2}x^{m}}{\partial X^{i}\partial X^{j}}\frac{\partial X^{k}}{\partial x^{m}}%
\text{ \ \ \ .}  \tag{A6}
\end{equation}%
Making use of the defining equation (A1) (but in terms of the initial
coordinates $x^{1},x^{2},....,x^{n}$), the above expression can be written
also as 
\begin{equation}
\widetilde{\Gamma }_{ij}^{k^{\prime }}(\mathbf{X})=\Gamma _{np}^{m}(\mathbf{x%
})\frac{\partial X^{k}}{\partial x^{\alpha }}\frac{\partial x^{n}}{\partial
X^{i}}\frac{\partial x^{p}}{\partial X^{j}}g_{m\beta }(\mathbf{x})dx^{\alpha
}dx^{\beta }+\frac{\partial ^{2}x^{\alpha }}{\partial X^{i}\partial X^{j}}%
\frac{\partial X^{k}}{\partial x^{\alpha }}\text{ \ \ \ .}  \tag{A7}
\end{equation}%
Clearly, if $\widetilde{\Gamma }_{ij}^{k^{\prime }}(\mathbf{X})$ is an
affine connection, from the R. H. S. of \ (A5) and (A7) it would follow that
the following relation has to be satisfied 
\begin{equation}
dx^{\alpha }dx^{\beta }g_{m\beta }(\mathbf{x})\frac{\partial ^{2}x^{m}}{%
\partial X^{i}\partial X^{j}}\frac{\partial X^{k}}{\partial x^{\alpha }}-%
\frac{\partial ^{2}x^{\alpha }}{\partial X^{i}\partial X^{j}}\frac{\partial
X^{k}}{\partial x^{\alpha }}=0\text{ \ \ ,}  \tag{A8}
\end{equation}%
which in fact represents the necesasy condition for the definition of the
connection $\widetilde{\Gamma }_{ij}^{k^{\prime }}(\mathbf{X})$ as an affine
connection. It can easily be proved that in case of commuting operators of
differentiation $\frac{\partial }{\partial x_{i}}$ and $\frac{\partial }{%
\partial X_{j}}$ (in the general case, however, they do not commute),
equation (A8) is fulfilled.

\subsubsection*{A1.2 THE TWO-DIMENSIONAL \ GENERALIZED \ CONNECTION \ $%
\widetilde{\Gamma }_{ij}^{k^{\prime }}$ IN \ THE \ GENERAL \ CASE}

\bigskip Now let us investigate the two-dimensional case, but when the
assumption about the commutation of the derivatives is dropped out. Then
equation (A8) on the integral curves $dx^{1}=C_{1}$ and $dx^{2}=C_{2}$ can
be written as (the relation $\frac{\partial x^{p}}{\partial X^{s}}\frac{%
\partial X^{s}}{\partial x^{t}}=\delta _{t}^{p}$ is also taken into account) 
\begin{equation*}
\frac{\partial X^{k}}{\partial x^{1}}\{(C_{1}^{2}g_{11}+C_{1}C_{2}g_{12}-1)%
\overset{(21)}{M^{lij}}-(C_{2}^{2}g_{22}+C_{1}C_{2}g_{12}-1)\overset{(12)}{%
M^{lij}}+
\end{equation*}%
\begin{equation}
+(C_{1}^{2}g_{12}+C_{1}C_{2}g_{22})\overset{(22)}{M^{lij}}%
-(C_{2}^{2}g_{12}+C_{1}C_{2}g_{11})\overset{(11)}{M^{lij}}\}=0\text{ ,} 
\tag{A9}
\end{equation}%
where $\overset{(kk)}{M^{lij}}$ and $\overset{(kn)}{M^{lij}}$ ($k,n=1$ or $2$%
) are the introduced notations for the expressions 
\begin{equation}
\overset{(kk)}{M^{lij}}:=\frac{\partial x^{k}}{\partial X^{l}}\frac{\partial
^{2}x^{k}}{\partial X^{i}\partial X^{j}}\text{ \ ; \ \ }\overset{(kn)}{%
M^{lij}}:=\frac{\partial x^{k}}{\partial X^{l}}\frac{\partial ^{2}x^{n}}{%
\partial X^{i}\partial X^{j}}\text{\ \ \ .}  \tag{A10}
\end{equation}%
Now interchanging the functions $x^{1}\leftrightarrow x^{2}$ in (A9) and
substracting the derived equation from (A9), one can obtain 
\begin{equation*}
\frac{\partial X^{k}}{\partial x^{1}}%
\{(C_{1}^{2}g_{11}-C_{2}^{2}g_{22})T^{lij}+
\end{equation*}%
\begin{equation}
+\left[ g_{12}(C_{1}^{2}+C_{2}^{2})+C_{1}C_{2}(g_{11}+g_{22})\right] (%
\overset{(22)}{M^{lij}}-\overset{(11)}{M^{lij}})\}=0\text{ \ \ \ ,} 
\tag{A11}
\end{equation}%
where $T^{lij}$ is an introduced notation for \ \ 
\begin{equation}
T^{lij}:=\frac{\partial x^{2}}{\partial X^{l}}\frac{\partial ^{2}x^{1}}{%
\partial X^{i}\partial X^{j}}-\frac{\partial x^{1}}{\partial X^{l}}\frac{%
\partial ^{2}x^{2}}{\partial X^{i}\partial X^{j}}\text{ \ .}  \tag{A12}
\end{equation}%
It can easily be checked that 
\begin{equation}
T^{[lij]}:=T^{lij}-T^{jil}=\frac{\partial }{\partial X^{i}}\left(
\{x^{1},x^{2}\}_{X^{j},X^{l}}\right) \text{ \ \ ,}  \tag{A13}
\end{equation}%
where $\{x^{1},x^{2}\}_{X^{j},X^{l}}$ is the notation for the s.c.
\textquotedblright one-dimensional Poisson bracket\textquotedblright \textbf{%
\ } 
\begin{equation}
\{x^{1},x^{2}\}_{X^{j},X^{l}}:=\frac{\partial x^{1}}{\partial X^{j}}\frac{%
\partial x^{2}}{\partial X^{l}}-\frac{\partial x^{1}}{\partial X^{l}}\frac{%
\partial x^{2}}{\partial X^{j}}\text{ \ \ .}  \tag{A14}
\end{equation}%
It can be proved that 
\begin{equation}
\overset{}{\overset{(kk)}{M^{(lij)}}=\frac{\partial }{\partial X^{i}}\left[ 
\frac{\partial x^{k}}{\partial X^{l}}\frac{\partial x^{k}}{\partial X^{j}}%
\right] }\text{ ;}\overset{(kk)}{M^{[lij]}=0}\text{ \ .}  \tag{A15}
\end{equation}%
Since we would like to obtain a relation by combining all the components of
eq. (A15) in the two-dimensional case, we can write down equation (A11) with
interchanged indices $l\leftrightarrow j$. Substracting the obtained
equation from (A11) and taking into account the \textquotedblright
antisymmetric\textquotedblright\ relations (A13)\ and (A15), one obtains the
simple equation 
\begin{equation}
\frac{\partial X^{k}}{\partial x^{1}}(C_{1}^{2}g_{11}-C_{2}^{2}g_{22})\frac{%
\partial }{\partial X^{i}}\{x^{1},x^{2}\}_{X^{j},X^{l}}=0\text{ \ \ \ .} 
\tag{A16}
\end{equation}%
Therefore in the general two-dimensional case of non-commuting operators of
differentiation, the \textquotedblright modified\textquotedblright\
connection $\widetilde{\Gamma }_{ij}^{k}$ has affine transformation
properties in each one of the following cases

1. If the generalized coordinates $X^{1}$ and $X^{2}$ do not depend on $%
x^{1} $.

2. If the Poisson bracket $\{x^{1},x^{2}\}_{X^{j},X^{l}}$ is constant on the
integral curves $dx^{1}=C_{1}$ and $dx^{2}=C_{2}.$

3. If the following relation is fulfilled for the metric components $g_{11}$
and $g_{22}$ and for the (arbitrary) constants $C_{1}$ and $C_{2}$ 
\begin{equation}
C_{1}^{2}g_{11}-C_{2}^{2}g_{22}=0\text{ \ \ .}  \tag{A17}
\end{equation}

\subsection*{A2: \ THE \ CONNECTION \ $\widetilde{\Gamma }_{ij}^{k}$ \ AS \
AN EQUIAFFINE \ CONNECTION}

\bigskip We have to prove that the connection $\widetilde{\Gamma }_{ij}^{k}$
for $j=k$ can be represented in the form of a gradient of a scalar quantity,
i. e. 
\begin{equation}
\widetilde{\Gamma }_{ij}^{k}=\partial _{i}lne\text{ \ \ .}  \tag{A18}
\end{equation}%
In the approximation ($dX^{i})_{,k}=0$ one can prove that the connection $%
\widetilde{\Gamma }_{ij}^{k}$ is indeed an equiaffine one, since one can set
up 
\begin{equation}
lne\equiv \frac{1}{2}dX^{k}dX^{s}g_{ks}\text{ \ \ \ .}  \tag{A19}
\end{equation}

The more complicated and interesting task is to prove that even in the case (%
$dX^{i})_{,k}\neq 0$, the connection $\widetilde{\Gamma }_{ij}^{k}$ will
again be an equiaffine one. For the purpose, note that 
\begin{equation}
\widetilde{\Gamma }_{ik}^{k}=\frac{1}{2}dX^{s}dX^{k}g_{ks,i}=\frac{1}{2}%
dX^{k}dX^{s}g_{r(s}\Gamma _{k)i}^{r}=W_{i}\text{ \ }  \tag{A20}
\end{equation}%
and consequently $\widetilde{\Gamma }_{ik}^{k}$ will be an equiaffine
connection if the scalar quantity $e$ can be determined as a solution of the
differential equation 
\begin{equation}
\partial _{i}lne=W_{i}\text{ \ }  \tag{A21}
\end{equation}%
as 
\begin{equation}
e=g(X_{1},X_{2},..,X_{i-1},X_{i+1},..,X_{n})e^{\int
W_{i}(X_{1},.....,X_{n})dX^{i}}\text{ \ .}  \tag{A22}
\end{equation}%
Note that the function $g$ depends on all variables $%
X_{1},X_{2},..,X_{i-1},X_{i+1},..,X_{n}$ with the exception of $X_{i}$,
while the function $W_{i}$ depends on all the variables, including also $%
X_{i}$.

Unfortunately, the proof at this stage will be incomplete, since $e$ will
depend on the choice of the variable $X_{i}$, which should not happen with a
scalar quantity. Consequently, it should be proved that the function $%
g(X_{1},X_{2},..,X_{i-1},X_{i+1},..,X_{n})$ can be determined in a proper
way (so that for every choice of $W_{i})$ the expression (A22) for $e$ would
be a scalar quantity. Until we have not proved it, we shall denote the L.H.
S. of (A22) with $e^{(i)}$.

Let us differentiate both sides of (A22) for $e\equiv e^{(i)}$ and $e\equiv
e^{(j)}$ by $X^{j}$ and $X^{i}$ respectively ($i\neq j$). We shall write
down only the first equation, since the second one is obtained from the
first after a change of the indices $i\Longleftrightarrow $ $j$. 
\begin{equation*}
\frac{\partial e^{(i)}}{\partial X^{j}}=\frac{\partial \ln
g(X_{1},X_{2},.,X_{i-1},X_{i+1},.,X_{n})}{\partial X^{j}}e^{(i)}+
\end{equation*}%
\begin{equation}
+g(X_{1},X_{2},..,X_{i-1},X_{i+1},..,X_{n})e^{\int \frac{\partial
W_{i}(X_{1},.....,X_{n})}{\partial X^{j}}dX^{i}}\text{ \ \ \ .}  \tag{A23}
\end{equation}%
Now differentiate again the derived equation (A23) for $\frac{\partial
e^{(i)}}{\partial X^{j}}$ by $X^{i}$ and the other equation for $\frac{%
\partial e^{(j)}}{\partial X^{i}}$ by $X^{j}$. Taking into account also that 
$\frac{\partial e^{(i)}}{\partial X^{i}}=e^{(i)}W_{i}$ , applying again
(A23) and defining summation over the indices $i$ and $j$, the result will
be 
\begin{equation*}
\sum_{i,j}\frac{\partial }{\partial X^{j}}\left( \frac{\partial e^{(i)}}{%
\partial X^{i}}\right) =grad\left[ \ln
g(X_{1},X_{2},.,X_{i-1},X_{i+1},.,X_{n})\right] (\mathbf{e}\text{ }.\mathbf{W%
})+
\end{equation*}%
\begin{equation}
+\sum_{i,j}\left[ \frac{\partial W_{i}}{\partial X^{j}}\frac{\partial e^{(i)}%
}{\partial X^{j}}-\frac{\partial \widetilde{W}_{ij}}{\partial X^{j}}.e^{(i)}%
\right] +(\mathbf{e}\text{ }.\mathbf{W})\bigtriangleup \ln
g(X_{1},X_{2},.,X_{i-1},X_{i+1},.,X_{n})\text{ \ \ \ \ ,}  \tag{A24}
\end{equation}%
where $(\mathbf{e}$ $.\mathbf{W})$ denotes a scalar product and the
following notation has been introduced

\begin{equation}
\widetilde{W}_{ij}\equiv W_{i}\frac{\partial \ln
g(X_{1},X_{2},.,X_{i-1},X_{i+1},.,X_{n})}{\partial X^{j}}\text{ \ \ \ .} 
\tag{A25}
\end{equation}%
Again, the second equation will be the same as (A24), but with $%
i\Leftrightarrow j$. Substracting the two equations and taking into account
the formulae for $graddive=\sum_{i,j}\frac{\partial }{\partial X^{i}}\left( 
\frac{\partial e^{(i)}}{\partial X^{j}}\right) $, one can derive 
\begin{equation*}
\sum_{i,j}\left[ \frac{\partial W_{i}}{\partial X^{j}}\frac{\partial e^{(i)}%
}{\partial X^{i}}-\frac{\partial W_{j}}{\partial X^{i}}\left( \frac{\partial
e^{(j)}}{\partial X^{i}}\right) \right] -
\end{equation*}%
\begin{equation}
-\sum_{i,j}\left[ \frac{\partial \widetilde{W}_{ij}}{\partial X^{j}}e^{(i)}-%
\frac{\partial \widetilde{W}_{ji}}{\partial X^{i}}e^{(j)}\right] =0\text{ \
\ \ \ .}  \tag{A26}
\end{equation}%
But it can be written also 
\begin{equation*}
\sum_{i,j}\left[ \frac{\partial W_{i}}{\partial X^{j}}\frac{\partial e^{(i)}%
}{\partial X^{i}}-\frac{\partial \widetilde{W}_{ij}}{\partial X^{j}}e^{(i)}%
\right] =
\end{equation*}%
\begin{equation}
=\sum_{i,j}\left[ \frac{\partial W_{i}}{\partial X^{j}}\frac{\partial e^{(i)}%
}{\partial X^{i}}-e^{(i)}\frac{\partial W_{i}}{\partial X^{j}}\frac{\partial
\ln g}{\partial X^{j}}-(\mathbf{e}\text{ .}\mathbf{W})\bigtriangleup \ln g%
\right] \text{ \ \ .}  \tag{A27}
\end{equation}%
Taking the above expression into account, equation (A26) acquires the form 
\begin{equation*}
\sum_{i,j}\left[ \frac{\partial W_{i}}{\partial X^{j}}\frac{\partial e^{(i)}%
}{\partial X^{i}}-\frac{\partial W_{i}}{\partial X^{j}}\frac{\partial e^{(i)}%
}{\partial X^{i}}\right] +
\end{equation*}%
\begin{equation}
+\sum_{i,j}\left[ e^{(j)}\frac{\partial W_{j}}{\partial X^{i}}\frac{\partial
\ln g}{\partial X^{i}}-e^{(i)}\frac{\partial W_{i}}{\partial X^{j}}\frac{%
\partial \ln g}{\partial X^{j}}\right] =0\text{ \ \ \ .}  \tag{A28}
\end{equation}%
Further we shall assume that each term in the sum is zero, i.e. the equation
is fulfilled for each $i$ and $j$. Substituting expressions (A22) for $e^{(i)%
\text{ }}$and $e^{(j)}$ and (A23) for $\frac{\partial e^{(i)}}{\partial X^{j}%
}$ and $\frac{\partial e^{(j)}}{\partial X^{i}}$, differentiating the
obtained expression by $X^{i}$ and making use again of (A22) and (A23), the
following simple differential equation can be obtained: 
\begin{equation*}
W_{j,i}\frac{\partial \ln g(X_{1},.,X_{j-1,}X_{j+1},..,X_{n})}{\partial X^{i}%
}+(W_{j,ii}-W_{i,j}W_{j,i}-
\end{equation*}%
\begin{equation}
-\frac{W_{i,ji}W_{j,i}}{W_{i,j}})+W_{j,i}e^{\int \left[ \frac{\partial
^{2}W_{j}}{\partial X^{i2}}-\frac{\partial W_{j}}{\partial X^{i}}\right]
dX^{j}}=0\text{ \ \ .}  \tag{A29}
\end{equation}%
The first case, when this equation will be satisfied will be 
\begin{equation}
W_{j,i}=\frac{\partial W_{j}}{\partial X^{i}}=0\text{ \ \ }\Longrightarrow
W_{j}=f(X_{1},..,X_{i-1},X_{i+1},..,X_{n})\text{ \ \ .}  \tag{A30}
\end{equation}%
Since this will be fulfilled for every $i$, then $W_{j}$ should be a
constant, which of course is a very rare and special case.

The second, more realistic case is when the function $g$ is a solution of
the differential equation (A23) for every $i$ and $j$ ($i\neq j$): 
\begin{equation}
g(X_{1},.,X_{j-1,}X_{j+1},..,X_{n})=F(X_{1},.,X_{i-1,}X_{i+1},..,X_{n})e^{%
\int \widetilde{Q}(X_{1},...,X_{n})dX^{i}}\text{ \ \ ,}  \tag{A31}
\end{equation}%
where 
\begin{equation}
\widetilde{Q}(X_{1},...,X_{n})\equiv \left( W_{i,j}+\frac{W_{i,ji}}{W_{i,j}}-%
\frac{W_{j,ii}}{W_{j,i}}\right) -e^{\int \left( \frac{\partial ^{2}W_{j}}{%
\partial X^{i2}}-\frac{\partial W_{j}}{\partial X^{i}}\right) dX^{j}}\text{ .%
}  \tag{A32}
\end{equation}%
Since the function $g(X_{1},.,X_{j-1,}X_{j+1},..,X_{n})$ on the L. H. S. of
(A31) does not depend on the variable $X_{j}$, then for each $j$ the unknown
function $F(X_{1},.,X_{i-1,}X_{i+1},..,X_{n})$ can be obtained after
differentiating both sides of (A31) by $X^{j}$. Thus the function $F$ is a
solution of the following differential equation 
\begin{equation}
0=\frac{\partial F(X_{1},.,X_{i-1,}X_{i+1},..,X_{n})}{\partial X^{j}}e^{\int 
\widetilde{Q}dX^{i}}+F(X_{1},.,X_{i-1,}X_{i+1},..,X_{n})e^{\int \frac{%
\partial \widetilde{Q}}{\partial X^{j}}dX^{i}}\text{ \ \ .}  \tag{A33}
\end{equation}%
This precludes the proof that the function $g$ in (A22) can be determined in
such a way that $e^{(i)}$ would be indeed a scalar quantity and therefore $%
e\equiv e^{(i)}$. Throughout the whole proof, we assumed that $W_{i}$,
determined by (A20), is a vector. This of course should be proved in the
same way, in which it was proved that the connection $\widetilde{\Gamma }%
_{ij}^{k}$ has affine transformation properties.

\end{document}